# Switching Properties in Magnetic Tunnel Junctions with Interfacial Perpendicular Anisotropy: Micromagnetic Study

R. Tomasello[1], V. Puliafito[2], B. Azzerboni[2], G. Finocchio[2]

[1]Department of Computer Science, Modelling, Electronics and System Science, University of Calabria, Via P. Bucci 87036, Rende (CS), Italy

[2]Department of Electronic Engineering, Industrial Chemistry and Engineering, University of Messina, 98166 Messina, Italy

The role of "universal memory" can be successfully satisfied by magnetic tunnel junctions (MTJs) where the writing mechanism is based on spin-transfer torque (STT). An improvement in the switching properties (lower switching current density maintaining the thermal stability) has been achieved in MTJs with interfacial perpendicular anisotropy (IPA) at the interface between CoFeB and MgO. In this paper, micromagnetic simulations point out the influence of IPA and saturation magnetization ($M_S$) on the properties of fast magnetization reversal achieved in 5, 10 and 20 ns. Both cases of in-plane and out-of-plane free layer are considered. In addition, the thermal effect is included for the in-plane switching at 20 ns and a complete analysis of energy dissipation during the switching is illustrated. This study can provide useful information for the design of STT-based memories.

*Index Terms*— micromagnetic model, interfacial perpendicular anisotropy, switching, MRAM

## I. INTRODUCTION

Magnetic Tunnel Junctions (MTJs) are becoming the first candidates to make non volatile spin transfer torque (STT) magnetoresistive random access memories (STT-MRAM) [1, 2]. The STT, together with an MgO tunnel barrier, has all the requirements to be the key ingredient for an STT-MRAM cell [3, 4, 5]. In fact, that kind of memory cell holds storage essential properties, such as high density, high speed, non-volatility and large endurance [6]. Nonetheless, in order to make such devices compatible with the "state of the art" CMOS technology, it is necessary to obtain high thermal stability and switching current density of the order of $10^6$ A/cm$^2$ for fast writing time (<20ns).

The magnetization reversal induced by spin torque was initially observed in in-plane structures [7, 8] and, therefore, it was successfully demonstrated in perpendicular spin valves [9] and MTJs [10, 11]. In such devices, it has been proved that an increase of the switching speed can be obtained with an out-of-plane polarizer and by modifying the anisotropy of the free layer. In particular, in structures with a CoFeB free layer and an MgO spacer, it is possible to exploit the interfacial perpendicular anisotropy (IPA) to achieve suitable switching [10] and dynamical properties [12].

The main advantage of a perpendicular device with respect to the in-plane one is the reduction of the critical current for magnetization reversal for an equal bit stability energy, $E = M_S V H_K / 2$, being $M_S$ the saturation magnetization, $V$ the free layer volume and $H_K$ the anisotropy field [10].

In detail, the critical current $I_C^{ip}$ necessary to achieve the in-plane switching can be expressed by:

$$I_C^{ip} \propto M_S V \left( H + H_K + 2\pi M_S - H_{K\perp} / 2 \right) \tag{1}$$

where $H$ is the external in-plane applied field, $2\pi M_S$ is related to the demagnetizing field and $H_{K\perp}$ is the perpendicular anisotropy field [13].

On the other hand, the critical current $I_C^{oop}$ for an out-of-plane switching is:

$$I_C^{oop} \propto M_S V \left( H + H_{K\perp} - 2\pi M_S \right) \tag{2}$$

being $2\pi M_S$ the demagnetizing field parallel to $H_{K\perp}$, and $H$ the perpendicular external field [13]. Those equations show that both the in-plane and the out-of-plane reversals depend on the perpendicular anisotropy and, thus, on the IPA.

IPA has been demonstrated to be due to an electrostatic interaction between the $2d$ iron orbital of the CoFeB ferromagnet with the $2p$ oxygen one of the MgO layer [10]. In order to model that interaction, it should be included as a magnetocrystalline field contribution $\mathbf{h}_{IPA}$ along the out-of-plane orientation and proportional to $K_u \left( \mathbf{m} \cdot \mathbf{u}_u \right)$, being $K_u$ the perpendicular anisotropy constant, $\mathbf{m}$ the magnetization vector and $\mathbf{u}_u$ the unit vector in the minimal anisotropy direction.

Experimental data in Fig. 1 of Ref. [12] show $M_S$ and IPA for a Co$_{20}$Fe$_{60}$B$_{20}$ free layer (FL) as function of its thickness, highlighting an inverse proportionality of the IPA. In the same paper, IPA is quantified by means of the product of $K_u$ and the FL thickness.

Here, we numerically study the dependence of the switching current density (SCD) on the IPA and on the $M_S$, either with an in-plane or with an out-of-plane magnetization, simulating a fast switching (20 ns) and an ultra-fast switching (10 and 5 ns). Moreover, we analyze the effect of the thermal contribution at room temperature, showing the critical current density which allows to reach a reliable magnetization reversal, as a function of the IPA and of $M_S$. We further provide a complete scheme of the power dissipation. With respect to the experimental works, where the IPA contribution is changed through the FL thickness or composition [14], our numerical analysis permits to determine directly the IPA value for the design of MRAMs.

## II. NUMERICAL DETAILS OF THE MICROMAGNETIC MODEL

In our numerical study, we reproduce the same experimental framework of Ref. [12]. The object under investigation is an MTJ nanopillar with an elliptical cross section of 170 x 60 nm$^2$, as displayed in Fig. 1. The sample structure stack includes PtMn(15) /Co$_{70}$Fe$_{30}$(2.3) /Ru(0.85) /Co$_{40}$Fe$_{40}$B$_{20}$(2.4) /MgO(0.8) /Co$_{20}$Fe$_{60}$B$_{20}$(1.4) (thicknesses in nm). The tunneling magnetoresistance (TMR) ratio and the resistance-area product are 102% and 3.8 $\Omega\mu m^2$, respectively. No external magnetic field is applied. We introduce a Cartesian coordinate system with the $x$-axis oriented along the larger dimension of the ellipse and the $z$-axis along the layers thickness. We consider the critical current to switch the magnetization from the antiparallel state to the parallel configuration.

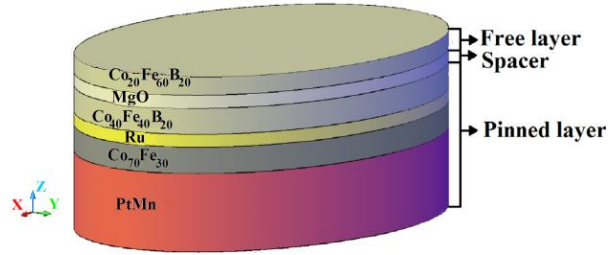

Fig. 1. Schematic representation of the studied MTJ.

We perform micromagnetic simulations based on the numerical solution of the Landau-Lifshitz-Gilbert-Slonczewski equation [15]:

$$\left(1+\alpha^2\right)\frac{d\mathbf{m}}{dt} = -\left(\mathbf{m}\times\mathbf{h}_{\text{eff}}\right) - \alpha\mathbf{m}\times\left(\mathbf{m}\times\mathbf{h}_{\text{eff}}\right)$$
$$-\frac{g\left|\mu_B\right|J}{e\gamma_0 M_s^2 d_{\text{FL}}}\varepsilon\left(\mathbf{m},\mathbf{m}_p\right)\left[\mathbf{m}\times\left(\mathbf{m}\times\mathbf{m}_P\right) - q\left(V\right)\left(\mathbf{m}\times\mathbf{m}_P\right)\right] \tag{3}$$

where $\alpha$ is the Gilbert damping, $\mathbf{m}$ is the magnetization vector, $\mathbf{h}_{\text{eff}}$ is the effective field, $g$ is the Landè factor, $\mu_B$ is the Bohr magneton, $e$ is the electron charge, $\gamma_0$ is the gyromagnetic ratio, $M_S$ is the saturation magnetization, $d_{\text{FL}}$ is the thickness of the free layer, $J$ is the current density, $\varepsilon(\mathbf{m},\mathbf{m}_p)$ characterizes the angular dependence of the Slonczewski spin torque term, $\mathbf{m}_P$ is the magnetization of the pinned layer, and $q(V)$ is a function of the applied voltage. We set $\alpha$=0.02 [16, 17]. Detailed numerical description of the model and algorithm can be found in Ref. [18, 19].

In order to consider an in-plane magnetization, we refer to the curves reported in Ref. [12] and we extract the FL thickness $d_{\text{FL}}$=1.4 nm, where the bulk anisotropy is null, and the corresponding $M_S$=8 x 10$^5$ A/m. After setting that value of the FL thickness, we study the in-plane switching properties (within a switching time close to 20 ns and to 10 ns) by changing the FL composition ($M_S$ and IPA). In practice, those parameters can be varied by changing the percentage of Co, Fe, and B. Particularly, it has been experimentally observed that the coercivity field (proportional to the term $H_{K\perp}-2\pi M_S$) changes with the free layer chemical composition [14] and, therefore, $M_S$ and IPA can be tuned in this way as well. Similarly, we analyze the out-of-plane fast switching with a switching time equal to 5 ns and 10 ns.

We also include the thermal effect (at $T$=350 K) for the specific case of in-plane fast magnetization reversal (<20ns). This effect on the magnetic system is modeled as an additive stochastic field added to the deterministic effective field in each computational cell, as explained in detail in a previous work [20].

## III. RESULTS AND DISCUSSION

Fig. 2 shows the SCD as function of the IPA and parameterized with three $M_S$ different values: 8 x 10$^5$ A/m, 10 x 10$^5$ A/m and 12 x 10$^5$A/m, as obtained by means of our numerical simulations.

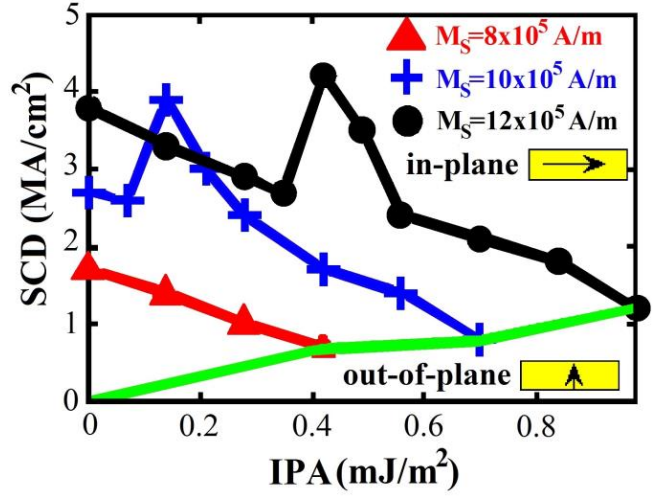

Fig. 2. In-plane switching current density depending on the perpendicular anisotropy. The three curves with symbols refer to a saturation magnetization equal to $8 \times 10^5$ A/m, $10 \times 10^5$ A/m and $12 \times 10^5$ A/m. The green line without symbols represents the minimum values of IPA over which, for a fixed $M_S$, the free layer is out-of-plane.

We can identify two regions in the phase diagram, divided by the solid line without symbols (green online): (i) the in-plane FL region where the magnetization is collinear to the polarizer; (ii) the out-of-plane region where the FL magnetization is perpendicular to the polarizer. The green line, in fact, has been determined from static simulations: for a fixed value of saturation magnetization, the point in the line represents the minimum IPA value which gives rise to the transition of the easy axis from in-plane to out-of-plane. This behavior is due to a competition between the IPA field and the demagnetizing field ($\mathbf{h}_d$). The latter exclusively depends on the geometrical properties of the FL, and, for an elliptical one, leads the magnetization to be in plane. Hence, until the $\mathbf{h}_{IPA}$ contribution is smaller than the $\mathbf{h}_d$ $z$-component, the magnetization keeps in-plane.

We observe that the in-plane SCD exhibits a non monotonic trend as a function of IPA for $M_S = 10 \times 10^5$ and $12 \times 10^5$ A/m. In fact, it is evident a first decrease, followed by a sudden increase till maximum points (peaks in the curves) are reached, for IPA equal to $1.4 \times 10^{-4}$ and to $4.2 \times 10^{-4}$ J/m$^2$, respectively. Then a monotonically decrease appears again, up to the IPA values where the FL magnetization turns out-of-plane, equal to $7 \times 10^{-4}$ and $9.8 \times 10^{-4}$ J/m$^2$, respectively. The SCD drops with the increase of the IPA because the in-plane magnetization becomes energetically less stable, making the magnetization reversal to be easier.

In addition, increasing the $M_S$ and keeping constant the IPA, the SCD increases as well. This is qualitatively [21] consistent with (1) and (3). In particular, in the latter equation, the $g \, |\mu_B| \, J / e \gamma_0 M_S^2 d_{FL}$ factor, that precedes the negative damping term, has to be maintained constant in order to balance the FL natural damping; thus, to an $M_S$ increase, an SCD rise has to correspond.

In order to explain the peaks in the curves, we have to refer to the spin domains excited in the device: when the IPA field is low, in the elliptical section, three magnetization domains are nucleated, which have a symmetric oscillation at the boundary, called edge mode (see Fig. 3).

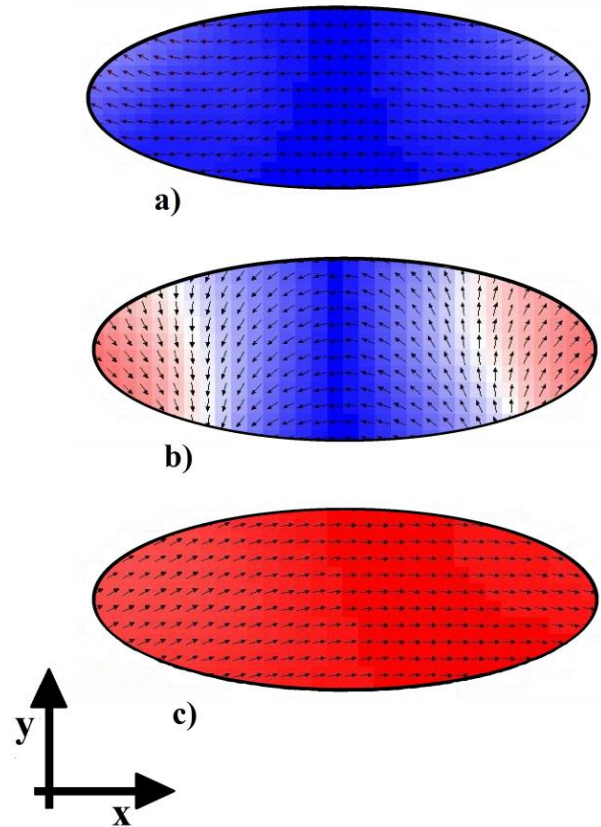

Fig. 3. Representation of the spin magnetization domains of the MTJ elliptical section, describing the in-plane switching phenomenon, when the spins are oriented along a) -x and c) x and b) during the switching.

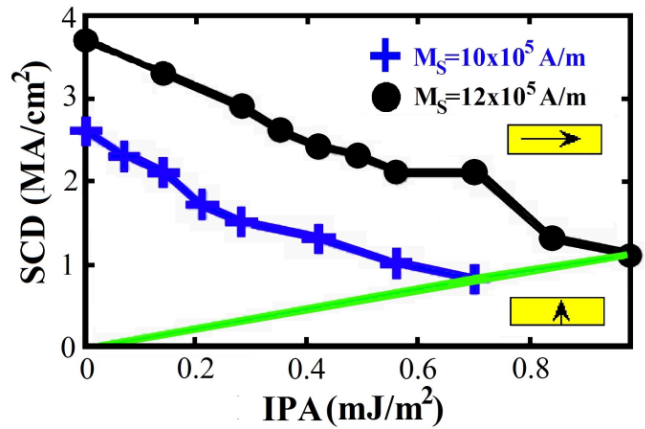

Fig. 4. In-plane switching current density as function of the perpendicular anisotropy, without the Oersted field. The two curves refer to a saturation magnetization equal to 10 x 10⁵ A/m and 12 x 10⁵A/m.

For the IPA values linked to the jumps in the curves (indicated above), the metastability of the domains is larger, due to a trade-off between the Oersted field and the STT [22, 23], leading to a longer switching for a fixed current. In fact, studying the switching without the Oersted field, the peaks disappear, as we see in Fig. 4. Moreover, we can underline that the observed decreasing linear relationship between the SCD and IPA in Fig. 4 qualitatively agrees with the analytical prediction of (1).

A similar study is carried out for a switching time of 10 ns. As we see in Fig. 5, a shorter switching time needs a higher SCD, about twice larger, whereas the SCD behavior as function of IPA and of $M_S$ is the same explained for a reversal time of 20 ns.

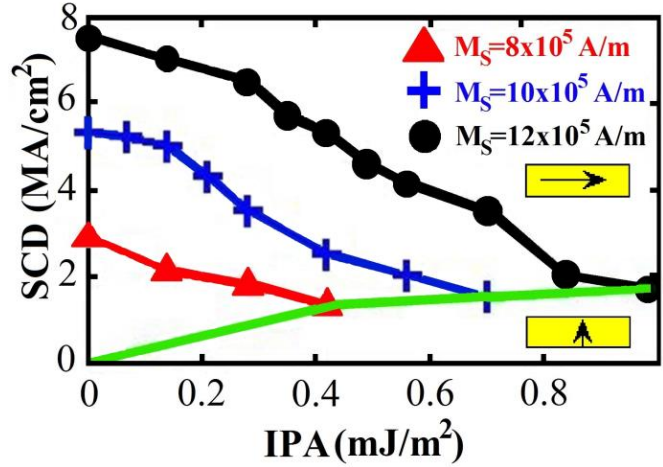

Fig. 5. In-plane switching current density depending on the perpendicular anisotropy, without the Oersted field and for a reversal time of 10 ns The three curves refer to a saturation magnetization equal to 8 x $10^5$ A/m, 10 x $10^5$ A/m and 12 x $10^5$ A/m.

For IPA values larger than 10 x $10^{-4}$ J/m$^2$, the FL magnetization is out-of-plane. In order to numerically simulate an out-of-plane structure, we orient the polarizer direction along the $z$-axis. Those results are illustrated in Fig. 6. Comparing the two graphs, we note a decrease of the SCD for longer switching time. Focusing only on Fig. 6a (the same evaluations can be extended to Fig. 6b), we show that the SCD increases linearly for higher IPA (as analytically confirmed by (2)), because of the larger stability of the magnetization $z$-component.

The SCD, for a fixed value of the IPA, increases with the reduction of the $M_S$. This is in agreement with the bulk anisotropy definition $K_b = K_U / t_{CoFeB} - 2\pi M_S$ [10, 12], where $t_{CoFeB}$ is the FL thickness. According to this relationship, higher $M_S$ values lead to a greater out-of-plane anisotropy contribution, favoring the energetic stability of the magnetization.

In addition, contrarily to the in-plane case, the out-of-plane results show a monotonic trend without peaks. This can be ascribed to the fact that the Oersted field has a null perpendicular component, so that it does not influence the out-of-plane SCD.

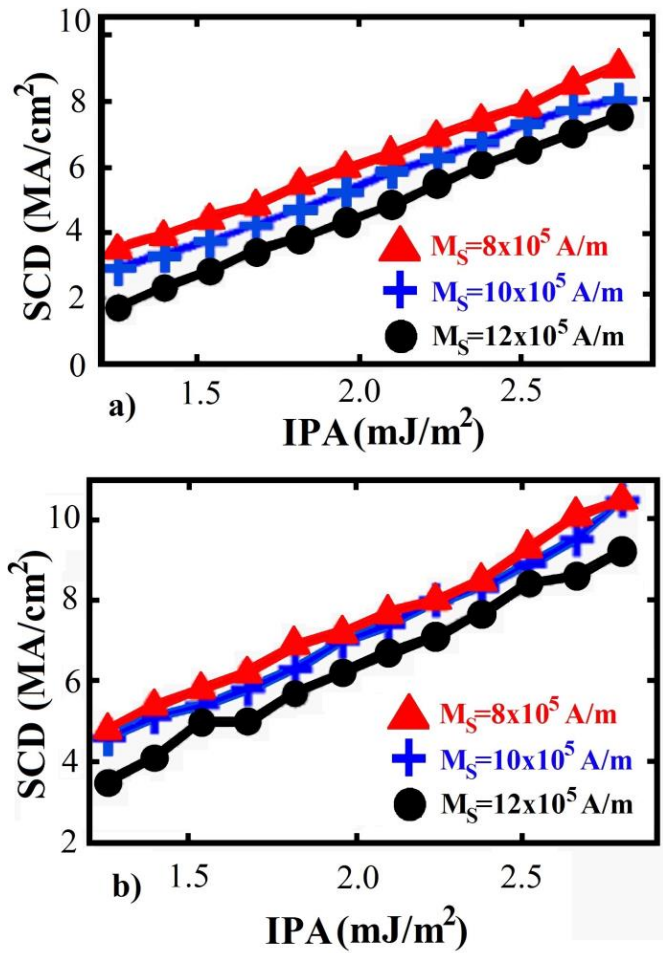

Fig. 6. Out-of-plane switching current density as function of the perpendicular anisotropy, for a switching time of a) 10 ns and b) 5 ns. The curves refer to a saturation magnetization equal to 8 x 10⁵ A/m, 10 x 10⁵ A/m and 12 x 10⁵ A/m.

Fig. 7 represents the SCD behavior as a function of IPA and of $M_S$ when the thermal contribution at room temperature $T$=350 K is considered and when the FL magnetization is in-plane, neglecting the Oersted field. The SCD plotted values are those ones that allow to obtain the switching phenomenon with a probability of 99.99 %, within a reversal time of 20 ns. As can be noted, the influences of IPA and of $M_S$ on the SCD are the same of the previous cases. Comparing those results with the results obtained for $T$=0 K (Figs. 2 and 4), we point out that the thermal contribution leads to a greater SCD reduction, up to 0.3 x 10⁶ A/cm² for IPA=4.2 x 10⁻⁴ J/m² and $M_S$=8 x 10⁵ A/m. These are very important results in term of applications, as the SCD values are compatible with the "state of art" semiconductor technology, and open interesting perspectives to make STT-MRAMs competitive.

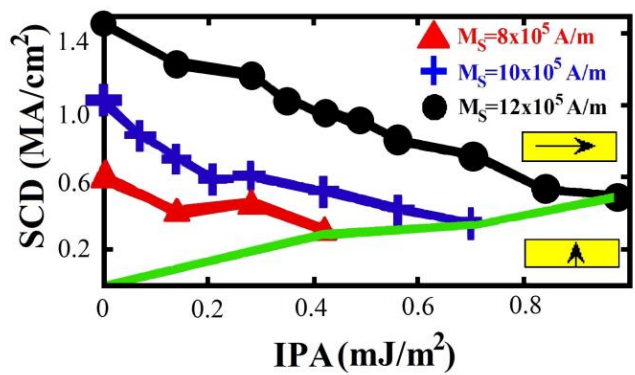

Fig. 7. In-plane switching current density as function of the perpendicular anisotropy, without the Oersted field and for a reversal time of 20 ns at a temperature $T$=350 K. The three curves refer to a saturation magnetization equal to 8 x 10⁵ A/m, 10 x 10⁵ A/m and 12 x 10⁵A/m.

A complete analysis of the device switching mean energy is shown in Fig. 8. The energy is computed by the following

expression [24]:

$$E = \sum \left( dt \cdot \left( J_S \cdot S \right)^2 \cdot R_p + \Delta R \left( \frac{1 - \cos \theta}{2} \right) \right) \qquad (4)$$

where $dt$ is the time step, $J_S$ is the SCD, $S$ is the elliptical cross section, $\theta$ is the angle between the magnetization of the FL and the polarizer, $R_p$ and $R_{ap}$ are the values of the resistance in the parallel and antiparallel states respectively and $\Delta R = R_p - R_{ap}$ (these values are extracted from the experimental data of the Ref. [12]). About the in-plane switching (Fig. 8a), by increasing the IPA, a reduction of the averaged energy is observed, but, at the same time, a worse thermal stability is achieved as well. Thus, in order to consider a device with good performance, it is necessary to find the right trade-off between the switching energy and the thermal stability limit value. As expected, studying the out-of-plane switching averaged energy (Figs. 8b and c), the lowest dissipated power is reached for the lowest IPA values and for a switching time of 10 ns.

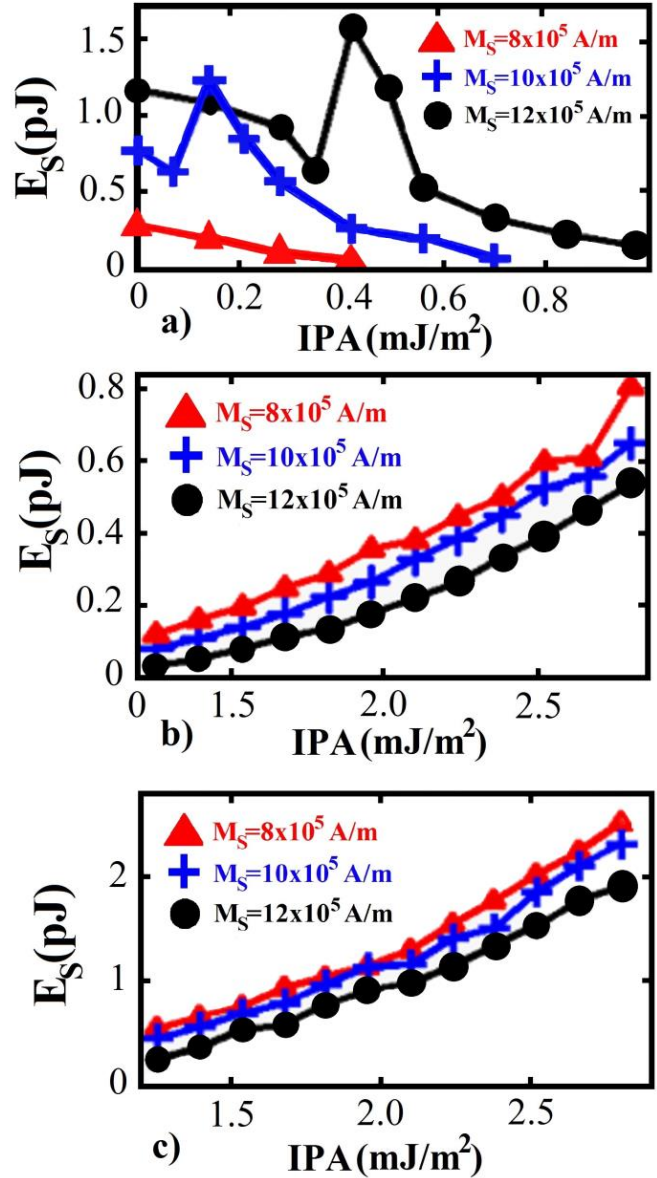

Fig. 8. Switching averaged energy as function of the IPA for a) the in-plane switching, b) out-of-plane switching in 10 ns and c) out-of-plane switching in 5 ns.

## IV. Conclusions

In this work we numerically studied the in-plane and out-of-plane SCD depending on the IPA and on $M_S$. We showed that the presence of the IPA leads an in-plane SCD decrease up to values close to $10^6$ A/cm$^2$, for all considered $M_S$ values and for a writing time of 20 ns. The switching averaged energy diminishes too, with a consequent smaller thermal stability. The effect of the thermal contribution at room temperature was the reduction of the SCD with respect to the SCD values at 0 K. On the contrary, about the out-of-plane switching, the lowest SCD values are reached for low IPA contributions and high $M_S$ ones, either for a writing time of 5 ns or 10 ns. In addition, for shorter writing time, a larger SCD is necessary and this means also a higher switching averaged energy. Finally, these achievements can be useful in the preliminary design stage of STT-MRAMs, as they provide interesting information on the switching currents and the corresponding energy consumption.

## ACKNOWLEDGEMENT

This work was supported by the project PRIN2010ECA8P3 from Italian MIUR.